\documentclass[showpacs,aps,prl]{revtex4}
\usepackage{graphicx}

\begin{document}
\title{Additive $N$-Step Markov Chains as Prototype Model of
 Symbolic Stochastic Dynamical Systems with Long-Range Correlations}

\author{Z.A. Mayzelis, S. S. Apostolov}

\affiliation{Department of Physics, Kharkov National University, 4
Svoboda Sq., Kharkov 61077, Ukraine}

\author{S. S. Melnyk, O. V. Usatenko
\footnote[1]{usatenko@ire.kharkov.ua},
 V. A. Yampol'skii}
\address{A. Ya. Usikov Institute for Radiophysics and Electronics \\
Ukrainian Academy of Science, 12 Proskura Street, 61085 Kharkov,
Ukraine}

\begin{abstract}
A theory of symbolic dynamic systems with long-range correlations
based on the consideration of the binary $N$-step Markov chains
developed earlier in Phys. Rev. Lett. \textbf{90}, 110601 (2003)
is generalized to the biased case (non equal numbers of zeros and
unities in the chain). In the model, the conditional probability
that the $i$-th symbol in the chain equals zero (or unity) is a
linear function of the number of unities (zeros) among the
preceding $N$ symbols. The correlation and distribution functions
as well as the variance of number of symbols in the words of
arbitrary length $L$ are obtained analytically and verified by
numerical simulations. A self-similarity of the studied stochastic
process is revealed and the similarity group transformation of the
chain parameters is presented. The diffusion Fokker-Planck
equation governing the distribution function of the $L$-words is
explored. If the persistent correlations are not extremely strong,
the distribution function is shown to be the Gaussian with the
variance being nonlinearly dependent on $L$. An equation
connecting the memory and correlation function of the additive
Markov chain is presented. This equation allows reconstructing a
memory function using a correlation function of the system.
Effectiveness and robustness of the proposed method is
demonstrated by simple model examples. Memory functions of
concrete coarse-grained literary texts are found and their
universal power-law behavior at long distances is revealed.
\end{abstract}

\date{\today}


\maketitle

\section{Introduction}

The problem of systems with long-range spatial and/or temporal
correlations (LRCS) is one of the topics of intensive research in
modern physics, as well as in the theory of dynamic systems and the
theory of probability. The LRC systems are usually characterized by
a complex structure and contain a number of hierarchic objects as
their subsystems. The LRCS are the subject of study in physics,
biology, economics, linguistics, sociology, geography, psychology,
etc.~\cite{stan,prov,mant,kant}.

One of the efficient methods to investigate the correlated systems
is based on a decomposition of the space of states into a finite
number of parts labeled by definite symbols. This procedure referred
to as coarse graining can be accompanied by the loss of short-range
memory between states of system but does not affect and does not
damage its robust invariant statistical properties on the large
scales. The most frequently used method of the decomposition is
based on the introduction of two parts of the phase space. In other
words, it consists in mapping the two parts of states onto two
symbols, say 0 and 1. Thus, the problem is reduced to investigating
the statistical properties of the symbolic binary sequences. This
method is applicable for the examination of both discrete and
continuous systems.

One of the ways to get a correct insight into the nature of
correlations consists in an ability of constructing a mathematical
object (for example, a correlated sequence of symbols) possessing
the same statistical properties as the initial system. There are
many algorithms to generate long-range correlated sequences: the
inverse Fourier transform~\cite{czir}, the expansion-modification Li
method~\cite{li}, the Voss procedure of consequent random
addition~\cite{voss}, the correlated Levy walks~\cite{shl},
etc.~\cite{czir}. We believe that, among the above-mentioned
methods, using the Markov chains is one of the most important.  This
was demonstrated in Ref.~\cite{uyakm}, where the Markov chains with
the \emph{step-like memory function} (MF) were studied. It was shown
that there exist some dynamical systems (coarse-grained sequences of
the Eukarya's DNA and dictionaries) with correlation properties that
can be properly described by this model.

The many-step Markov chain is the sequence of symbols of some
alphabet constructed using a conditional probability function,
which determines the probability of occurring some definite symbol
of sequence depending on $N$ previous ones. The property of
additivity of Markov chain means the \emph{independent} influence
of different previous symbols on generated one. The concept of
additivity, primarily introduced in paper~\cite{uya}, was later
generalized for the case of binary \emph{non-stationary} Markov
chains~\cite{hod}. Another generalization was based on
consideration of Markov sequences with a many-valued
alphabet~\cite{nar1, nar2}. Here we generalize the results of
paper~\cite{uyakm} to the biased case where the numbers of zeros
and unities are not supposed to be equal.

In the present work, we also continue investigating into additive
Markov chains with more complex memory functions. An equation
connecting mutually-complementary characteristics of a random
sequence, i.e. the memory and correlation functions, is obtained.
Upon finding the memory function of the original random sequence
on the basis of the analysis of its statistical properties,
namely, its correlation function, we can build the corresponding
Markov chain, which possesses the same statistical properties as
the initial sequence.

\section{Formulation of the problem}

\subsection{Conditional probability of the many-step
additive Markov chain}

Let us consider a stationary binary sequence of symbols $a_{i}$,
$a_{i}=\{0,1\}$, $i\in \textbf{\textbf{Z}} =...,-1,-2,0,1,2,...$.
To determine the $N$-\textit{step Markov chain} we have to
introduce the \emph{conditional probability} $P(a_{i}\mid
a_{i-N},a_{i-N+1},\dots ,a_{i-1})$ of occurring the definite
symbol $a_i$ (for example, $a_i =1$) after $N$-word $T_{N,i}$,
where $T_{N,i}$ stands for the sequence of symbols
$a_{i-N},a_{i-N+1},\dots ,a_{i-1}$. Thus, it is necessary to
define $2^{N}$ values of the $P$-function corresponding to each
possible configuration of the symbols $a_i$ in the $N$-word
$T_{N,i}$.

We suppose that the conditional probability $P(a_{i}\mid T_{N,i})$
differs from zero and unity for any word $T_{N,i}$ that provides
the metrical transitivity of the Markov chain (see Appendix). In
turn, according the Markov theorem, this property leads to the
ergodicity of the symbolic system under consideration.

Since we suppose to apply our theory to the sequences with long
memory lengths of the order of $10^6$, some special restrictions
to the class of $P$-functions should be imposed. We consider the
memory function of the \textit{additive} form,
\begin{equation}
P(a_{i}=1\mid T_{N,i}) =
\frac{1}{N}\sum\limits_{r=1}^{N}f(a_{i-r},r). \label{1}
\end{equation}
Here the function $f(a_{i-k},k)/N$ describes the additive
contribution of the symbol $a_{i-r}$ to the conditional
probability of occurring the symbol unity, $a_{i}=1$, at the $i$th
site. The homogeneity of the Markov chain is provided by
independence of the conditional probability Eq.~(\ref{1}) of the
index $i$. It is possible to consider Eq.~(\ref{1}) as the first
term in expansion of conditional probability in the formal series,
where each term corresponds to the additive (unary), binary,
ternary, and so on functions up to the $N$-ary one.

It is reasonable to assume the function $f$ to be decreasing with
an increase of the distance $r$ between the symbols $a_{i-r}$ and
$a_{i}$ in the Markov chain. However, for the sake of simplicity
we consider a step-like memory function $f(a_{i-r},r)$ independent
of the second argument $r$. As a result, the model is
characterized by three parameters only, specifically by $f(0)$,
$f(1)$, and $N$:
\begin{equation}
P(a_{i}=1\mid T_{N,i}) = \frac{1}{N}\sum\limits_{r=1}^{N}f(a_{i-r}).
\label{2}
\end{equation}
Note that the probability $P$ in Eq.~(\ref{2}) depends on the
numbers of symbols 0 and 1 in the $N$-word but is independent of the
arrangement of the elements $a_{i-k}$. Instead of two parameters
$f(0)$ and $f(1)$ it is convenient to introduce new independent
parameters $\nu$ and $\mu$ (see below Eq.~(\ref{3})),
\begin{equation}
f(0)+f(1)=1+2\nu, \qquad |\nu| <1/2.  \label{2a}
\end{equation}
Parameter $\nu$ provides the statistical inequality of the numbers
of symbols zero and unity in the Markov chain under consideration.
In other words, the chain is biased. Indeed, taking into account
Eqs.~(\ref{2}) and (\ref{2a}) and the sequence of equations,
\begin{equation}
P(a_{i} =
1|T_{N,i})=\frac{1}{N}\sum\limits_{r=1}^{N}f(\tilde{a}_{i-r})-2\nu =
P(a_{i}=0\mid \tilde{T}_{N,i})-2\nu, \label{2b}
\end{equation}
one can see the lack of symmetry with respect to interchange
$\tilde{a}_{i}\leftrightarrow a_{i}$ in the Markov chain if
$\nu\neq 0$. Here $\tilde{a}_{i}$ is the symbol ''opposite'' to
$a_{i}$, $\tilde{a}_{i}=1-a_{i}$, and $\tilde{T}_{N,i}$ is the
word ''opposite'' to ${T}_{N,i}$. Therefore, the probabilities of
occurring the words ${T}_{L,i}$ and $\tilde{T}_{L,i}$ are not
equal to each other for any word of the length $L$. At $L=1$ this
yields nonequal average probabilities that symbols $0$ and $1$
occur in the chain. Particularly, probability of occurring symbol
$0$ is grater by $2\nu$ than that of symbol $1$. If $\nu=0$ one
has non-biased case.

Taking into account the symmetry of the conditional probability
$P$ with respect to a permutation of symbols $a_{i}$ (see
Eq.~(\ref{2})), we can simplify the notations and introduce the
conditional probability $p_{k}$ of occurring the symbol zero after
the $N$-word containing $k$ unities, e.g., after the word
$\underbrace{(11...1}_{k}\;\underbrace{00...0}_{N-k})$,
\[
p_{k}=P(a_{N+1}=0\mid \underbrace{11\dots
1}_{k}\;\underbrace{00\dots 0} _{N-k})
\]
\begin{equation}
=\frac{1}{2}+\nu+\mu \left(1-\frac{2k}{N}\right),  \label{14}
\end{equation}
with the correlation parameter $\mu $ being defined by the
relation
\begin{equation}
\mu =\frac{f(0)-f(1)}{2}=f(0)-\frac{1}{2}-\nu.  \label{3}
\end{equation}

We focus  mainly our attention on the region of $\mu $ determined
by the persistence inequality $0 < \mu $. In this case, each of
the symbols unity in the preceding $N$-word promotes the birth of
new symbol unity. Nevertheless, the major part of our results is
valid for the anti-persistent region $\mu <0$ as well. Note that
inequalities $|\nu|<1/2$ and $|\mu+\nu|<1/2$ follow from
Eq.~(\ref{14})). Without loss of generality, we consider a case
$\nu>0$ only.

\subsection{Statistical characteristics of the chain}

In order to investigate the statistical properties of the Markov
chain, we consider the distribution $W_{L}(k)$ of the words of
definite length $L$ by the number $k$ of unities in them,
\begin{equation}
k_{i}(L)=\sum\limits_{l=1}^{L}a_{i+l},  \label{5}
\end{equation}
and the variance of $k$,
\begin{equation}
D(L)=\overline{k^{2}}-\overline{k}^{2},  \label{7}
\end{equation}
where
\begin{equation}
\overline{g(k)}=\sum\limits_{k=0}^{L}g(k)W_{L}(k).  \label{8}
\end{equation}
If $\mu =0,$ one arrives at the known result for the
non-correlated Brownian diffusion,
\begin{equation}
D(L)=L\left(\frac{1}{4}-\nu^2\right).  \label{6}
\end{equation}
We will show that the distribution function $W_{L}(k)$ for the
sequence determined by Eq.~(\ref{14}) (with nonzero but not
extremely close to $1/2-\nu$ parameter $\mu $) is the Gaussian
with the variance $D(L)$ nonlinearly dependent on $L$. However, at
$\mu \rightarrow 1/2 -\nu$ the distribution function can differ
strongly from the Gaussian.

\subsection{Main equation}

For the stationary Markov chain, the probability
$b(a_{1}a_{2}\dots a_{N})$ of occurring a certain word
$(a_{1},a_{2},\dots ,a_{N})$ satisfies the condition of
compatibility for the Chapman-Kolmogorov equation (see, for
example, Ref.~\cite{gar}):
\[
b(a_{1}\dots a_{N})=
\]
\begin{equation}
=\sum_{a=0,1}b(aa_{1}\dots a_{N-1})P(a_{N}\mid a,a_{1},\dots
,a_{N-1}).  \label{10}
\end{equation}
Thus, we have $2^{N}$ homogeneous algebraic equations for the
$2^{N}$ probabilities $b$ of occurring the $N$-words and the
normalization equation $\sum b=1$. This set of equations is
equivalent to that of Eq.~(\ref{eq8}). In the case under
consideration, the set of Eqs.~(\ref{10}) can be substantially
simplified owing to the following statement:

\textbf{Proposition 1}: \textit{The probability
$b(a_{1}a_{2}\dots a_{N})$ depends on the number $k$ of
unities in the $N$-word only}, i.\ e., it is independent of
the arrangement of symbols in the word $(a_{1},a_{2},\dots
,a_{N})$.
\begin{figure}[h!]
{\includegraphics[width=0.48\textwidth,height=0.32\textwidth]
{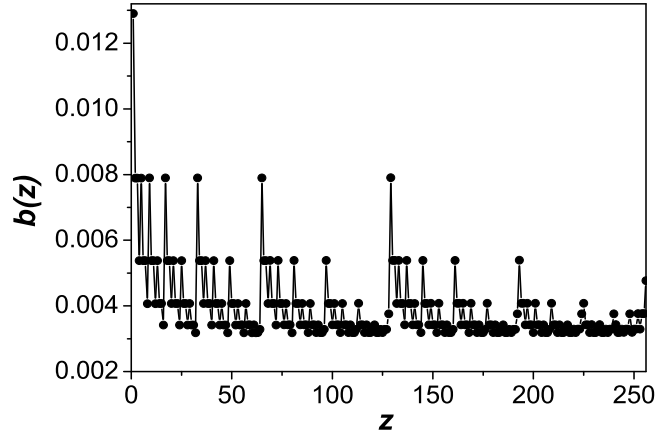}} \caption{The probability $b$ of occurring a word $(a_1,
a_2, \dots , a_ N)$ vs its number $z$ expressed in the binary code,
$z=\sum_{i=1}^N a_i \cdot 2^{i-1}$, for $N=8$, $\mu=0.1$, $
\nu=0.03$.} \label{f1}
\end{figure}

This statement illustrated by Fig.~1 is valid owing to the chosen
simple model (\ref{2}), (\ref{14}) of the Markov chain. It can be
easily verified directly by substituting the obtained below
solution (\ref{b}) into the set of Eqs.~(\ref{10}). Note that
according to the Markov theorem, Eqs.~(\ref{10}) do not have other
solutions~\cite{kat}.

Proposition 1 evidently leads to the very important property of
isotropy: any word $(a_{1},a_{2},\dots ,a_{L})$ appears with the
same probability as the inverted one, $(a_{L},a_{L-1},\dots
,a_{1})$.

Let us apply the set of Eqs.~(\ref{10}) to the word
$(\underbrace{11\dots 1}_{k}\;\underbrace{00\dots
0}_{N-k})$:
\[
b(\underbrace{11\dots 1}_{k}\;\underbrace{00\dots 0}_{N-k}) =b(0
\underbrace{11\dots 1}_{k}\;\underbrace{00\dots 0}_{N-k-1})p_{k}+
\]
\begin{equation} \label{13}
+b(1\underbrace{11\dots 1}_{k}\;\underbrace{00\dots
0}_{N-k-1})p_{k+1}.
\end{equation}
This yields the recursion relation for
$b(k)=b(\underbrace{11...1}_{k}\; \underbrace{00...0}_{N-k})$,
\[
b(k)=\frac{1-p_{k-1}}{p_{k}}b(k-1)=
\]
\begin{equation}
=\frac{N-2\nu N-2\mu (N-2k+2)}{N+2\nu N+2\mu (N-2k)} b(k-1).\label{15}
\end{equation}
The probabilities $b(k)$ for $\mu>0$ satisfy the sequence of
inequalities,

\[
b\left(\frac{N}{2}\left(1+\frac{\nu}{\mu}\right)\right)<
b\left(\frac{N}{2}\left(1+\frac{\nu}{\mu}\right)-1\right)<...<b(0),
\]
\begin{equation}
b\left(\frac{N}{2}\left(1+\frac{\nu}{\mu}\right)\right)<
b\left(\frac{N}{2}\left(1+\frac{\nu}{\mu}\right)+1\right)<...<b(N),
\label{15b}
\end{equation}
which is the reflection of persistent properties for the chain.

The solution of Eq.~(\ref{10}) is
\begin{equation}\label{b}
b(k)=A\cdot \Gamma ( n_1+k) \Gamma ( n_2+N-k)
\end{equation}
with the parameters $n_1$ and $n_2$ defined by
\begin{equation}\label{18a}
n_1= \frac{N(1-2(\mu+\nu))}{4\mu},\ \ \ n_2=
\frac{N(1-2(\mu-\nu))}{4\mu} .
\end{equation}
The constant $A$ will be found below by normalizing the
distribution function. Its value is,
\begin{equation}
A=\frac{\Gamma(n_1+n_2)}{\Gamma(n_1)\Gamma(n_2)\Gamma(n_1+n_2+N)}.\label{17a}
\end{equation}

\section{Distribution function of $L$-words}

In this section we investigate into the statistical properties of
the Markov chain, specifically, the distribution of the words of
definite length $L$ by the number $k$ of unities. The length $L$
can also be interpreted as the number of jumps of some particle
over an integer-valued 1D lattice or as the time of the diffusion
imposed by the Markov chain under consideration. The form of the
distribution function $W_{L}(k)$ depends, to a large extent, on
the relation between the word length $L$ and the memory length
$N$. Therefore, the first thing we will do is to examine the
simplest case $L = N$.

\subsection{Statistics of $N$-words}

The value $b(k)$ is the probability that an $N$-word contains $k$
unities with a \textit{definite} order of symbols $a_i$.
Therefore, the probability $W_{N}(k)$ that an $N$-word contains
$k$ unities with \textit{arbitrary} order of symbols $a_i$ is
$b(k)$ multiplied by the number $\mathrm{C}_{N}^{k}=N!/k!(N-k)!$
of different permutations of $k$ unities in the $N$-word,
\begin{equation}
W_{N}(k)=\text{C}_{N}^{k}b(k).  \label{19}
\end{equation}
Combining Eqs.~(\ref{b}) and (\ref{19}), we find the
distribution function,
\begin{equation}
W_{N}(k)= W_{N}(0)\text{C}_{N}^{k}\frac{\Gamma ( n_1+k) \Gamma (
n_2+N-k) }{\Gamma (n_1 ) \Gamma (n_2+N)}.  \label{18}
\end{equation}
The normalization constant $W_{N}(0)$ can be obtained from the
equality $\sum\limits_{k=0}^{N}W_{N}(k)=1$,
\begin{equation}
W_N(0)=\frac{\Gamma(n_1+n_2)\Gamma(n_2+N)}{\Gamma(n_2)\Gamma(n_1+n_2+N)}.\label{17}
\end{equation}
Comparing Eqs.~(\ref{b}), (\ref{19})-(\ref{17}), one can get
Eq.~(\ref{17a}) for the constant $A$ in Eq.~(\ref{b}).

\subsubsection{Limiting case of weak persistence, $n_1, n_2 \gg 1$}

In terms of the correlation parameter $\mu$, this limiting
case corresponds to the values of $\mu$ not very close to
1/2,
\begin{equation}\label{mu-wp}
 \frac{1-2(\mu +\nu)}{4\mu}\gg \frac{1}{N}.
\end{equation}
This inequality can be rewritten via the $f$-function (see
Eqs.~(\ref{2})---(\ref{3})),
\begin{equation}\label{f-wp}
\frac{f(1)}{f(0)-f(1)}\gg \frac{1}{N}.
\end{equation}

In the absence of correlations, $n_2 \rightarrow \infty$,
Eq.~(\ref{18}) and the Stirling formula yield the Gaussian
distribution at $k,\, N n_1/n_2,\, N-k \gg 1$, $k-k_0\ll N$. Given
the persistence is not too strong,
\begin{equation}\label{19c}
n_2 \gg 1,
\end{equation}
one can also obtain the Gaussian form for the distribution
function,
\begin{equation}
W_{N}(k)=\frac{1}{\sqrt{2\pi D(N)}}\exp \left\{
-\frac{(k-k_0)^{2}}{2D(N)} \right\} ,  \label{27}
\end{equation}
with the $\mu$-dependent variance,
\[
D(N)=\frac{N(N+n_1+n_2)n_1 n_2}{(n_1+n_2)^3}=
\]
\begin{equation}
=\frac{N}{4(1-2\mu )}\left[1-\frac{4\nu^2}{(1-2\mu)^2}\right],
\label{28}
\end{equation}
\begin{equation}
k_0=\frac{n_1}{n_1+n_2}N=\frac{N}{2(1-2\mu)}\left[1-\frac{2\nu}
{1-2\mu}\right].
\end{equation}
It is followed from Eq.~(\ref{27}) that $N$-words containing $k_0$
unties are the most probable. It is interesting to note, that the
persistence leads to a \emph{decrease} of the variance $D(N, \mu
>0)$ with respect to $D(N, \mu=0)=N\left(1/4-\nu^2\right)$ if
\begin{equation}\label{288a}
\nu>\frac{1-2\mu}{2\sqrt{3-6\mu+4\mu^2}}.
\end{equation}
In other case, for instance, at $\nu =0$, the persistence results
in an increase of the variance $D(N, \mu)$.  To put it
differently, the persistence is conductive to the intensification
of the diffusion under conditions opposite to inequality
(\ref{288a}).

Inequality $n_2 \gg 1$ gives $D(N) \ll N^{2}$. Therefore, despite
the increase of $D(N)$, the fluctuations of $(k-k_0)$ of the order
of $N$ are exponentially small.

\subsubsection{Intermediate case, $n_2 \gtrsim 1$}

If the parameters $n_1$ and $n_2$ are integers of the order of
unity, the distribution function $W_{N}(k)$ is a polynomial of
degree $n_1+n_2-2$. In particular, at $n_1=n_2=1$, the function
$W_{N}(k)$ is constant,
\begin{equation}
W_{N}(k)=\frac{1}{N+1}.  \label{24}
\end{equation}
At $n_1\neq 1$, $W_{N}(k)$ has a maximum within the interval
$[0,N]$. At $n_1= 1$ and $n_2>1$, $W_{N}(k)$ decreases
monotonously with an increase of $k$.

\subsubsection{Limiting case of strong persistence}
If the parameter $n_2$ satisfies the inequality,
\begin{equation}\label{24a}
n_2 \ll \ln^{-1}N,
\end{equation}
or
\begin{equation}\label{mu-sp}
  1-2(\mu-\nu) \ll 1/N\ln(N), \qquad f(1) \ll 1/N\ln(N),
\end{equation}
then one can neglect the parameters $n_1$ and $n_2$ in the
arguments of the functions $\Gamma (n_1+k)$, $\Gamma (n_2+N)$, and
$\Gamma (n_2+N-k)$ in Eq.~(\ref{18}). In this case, the
distribution function $W_{N}(k)$ assumes its maximal values at
$k=0$ and $k=N$,
\begin{equation}
W_{N}(1)=W_{N}(0)\frac{n_1 N}{N-1} \ll W_{N}(0). \label{20}
\end{equation}
Formula (\ref{20}) describes the sharply decreasing
$W_{N}(k)$ as $k$ varies from $0$ to $1$ (and from $N$ to
$N-1$). Then, at $1<k<N/2$, the function $W_{N}(k)$
decreases more slowly with an increase in $k$,
\begin{equation}
W_{N}(k)=W_{N}(0)\frac{n_1 N}{k(N-k)}.  \label{21}
\end{equation}
At $k=N/2,$ the probability $W_{N}(k)$ achieves its minimal value,
\begin{equation}
W_{N}\left(\frac{N}{2}\right)= W_{N}(0)\frac{4n_1}{N}. \label{22}
\end{equation}

It follows from normalization (\ref{17}) that the values
$W_{N}(0)$ and $W_N (N)$ are approximatively equal to
$n_2/(n_1+n_2)$ and $n_1/(n_1+n_2)$ respectively. Neglecting the
terms of the order of $n_2^2$, one gets
\begin{equation}
W_{N}(0)=\frac{n_2}{n_1+n_2} ( 1 - n_1 \ln N ), \label{22a1}
\end{equation}
\begin{equation}
W_{N}(N)=\frac{n_1}{n_1+n_2} ( 1 - n_2 \ln N ). \label{22a2}
\end{equation}
In the straightforward calculation using Eqs. (\ref{7}) and
(\ref{21}) the variance $D$ is
\begin{equation}
D(N)=\frac{n_1 n_2 N^2}{(n_1+n_2)^2} -\frac{n_1 n_2
N(N-1)}{n_1+n_2}. \label{22b}
\end{equation}

Thus, the variance $D(N)$ is equal to $n_1 n_2 N^2 /(n_1+n_2)$ in
the leading approximation in the parameter $n$. This fact has a
simple explanation. The probability of occurrence the $N$-word
containing $N$ unities is approximatively equal to
$n_1/(n_1+n_2)$. So, the relations $\overline{k^{2}} \approx n_1
N^2/(n_1+n_2) $ and $\overline{k}^{2}=n_1^2 N^2/(n_1+n_2)^2$ give
(\ref{22b}). The case of strong persistence corresponds to the
so-called ballistic regime of diffusion: if we chose randomly some
symbol $a_i$ in the sequence, it will be surrounded by the same
symbols with the probability close to unity.

The evolution of the distribution function $W_N(k)$ from the
Gaussian form to the inverse one with a decrease of the parameters
$n_1$ and $n_2$ is shown in Fig.~2. In the interval $\ln^{-1}N <
n_2 < 1 $ the curve $W_{N}(k)$ is concave and the maximum of
function $W_{N}(k)$ inverts into minimum. At $N \gg 1 $ and
$\ln^{-1}N < n_2 < 1 $, the curve remains a smooth function of its
argument $k$ as shown by curve with $n=0.5$ in Fig.~2. Below, we
will not consider this relatively narrow region of the change in
the parameter $n_2$.

Formulas (\ref{27}), (\ref{28}), (\ref{21}) and  (\ref{22a1}) ---
(\ref{22b}) describe the statistical properties of $L$-words for
the fixed ''diffusion time'' $L=N$. Below, we examine the
distribution function $W_{L}(k)$ for more general situation, $L<
N$.
\begin{figure}[h!]
{\includegraphics[width=0.45\textwidth,height=0.35\textwidth]{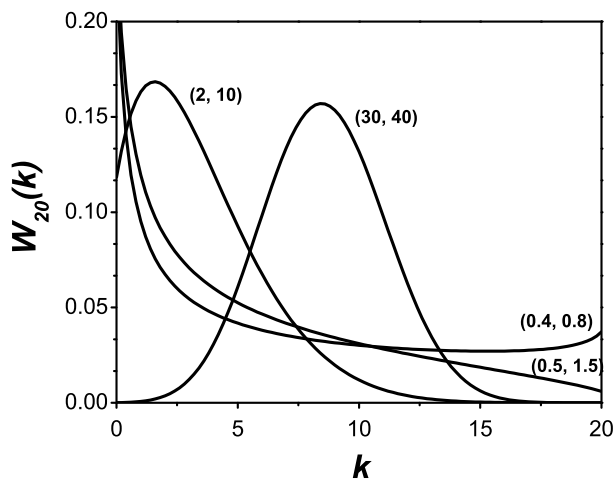}}
\caption{The distribution function $W_N(k)$ for $N$=20 and
different values of the parameters $n_1$ and $n_2$ shown near the
curves.} \label{f2}
\end{figure}
\subsection{Statistics of $L$-words with $L<N$}

\subsubsection{Distribution function $W_{L}(k)$}

The distribution function $W_{L}(k)$ at $L<N$ can be given as
\begin{equation}
W_{L}(k)=\sum\limits_{i=k}^{k+N-L}b(i)\text{C}_{L}^{k}\text{C}_{N-L}^{i-k}.
\label{29}
\end{equation}
This equation follows from the consideration of $N$-words
consisting of two parts,
\begin{equation}
(\underbrace{a_{1},\dots ,a_{N-L},}_{i-k\text{
unities}}\;\underbrace{a_{N-L+1},\dots ,a_{N}}_{k \text{
unities}}). \label{29b}
\end{equation}
The total number of unities in this word is $i$. The
right-hand part of the word ($L$-sub-word) contains $k$
unities. The remaining ($i-k$) unities are situated within
the left-hand part of the word (within $(N-L)$-sub-word).
The multiplier $\mathrm{C}_{L}^{k}\mathrm{C}_{N-L}^{i-k}$ in
Eq.~(\ref{29}) takes into account all possible permutations
of the symbols ''1'' within the $N$-word on condition that
the $L$-sub-word always contains $k$ unities. Then we
perform the summation over all possible values of the number
$i$. Note that Eq.~(\ref{29}) is a direct consequence of the
proposition 1 formulated in Subsec.~C of the previous
section.

The straightforward summation in Eq.~(\ref{29}) yields the
following formula that is valid at any value of the parameters
$n_1$ and $n_2$:
\begin{equation}\label{W(L)}
W_L(k)=W_L(0)\text{C}_{L}^{k}\frac{\Gamma(n_1+k) \Gamma
(n_2+L-k)}{\Gamma (n_1) \Gamma (n_2+L)}
\end{equation}
where
\begin{equation}\label{W(0)}
W_L(0)=\frac{\Gamma(n_1+n_2)\Gamma (n_2+L)}{\Gamma(n_2)\Gamma
(n_1+n_2+L)}.
\end{equation}

It is of interest to note that the parameters $\mu$, $\nu$ and the
memory length $N$ are presented in Eqs.~(\ref{W(L)}), (\ref{W(0)})
via the parameters $n_1$ and $n_2$ only. This means that the
statistical properties of the $L$-words with $L<N$ are defined by
these "combined" parameters.

In the limiting case of weak persistence, $n_2\gg 1$, at
$k,\;Ln_1/n_2,\;L-k \gg 1$, Eq.~(\ref{W(L)}) along with the
Stirling formula give the Gaussian distribution function,
\begin{equation}
W_{L}(k)=\frac{1}{\sqrt{2\pi D(L)}}\exp \left\{
-\frac{(k-k_0)^{2}}{2D(L)} \right\}  \label{31}
\end{equation}
with the variance $D(L)$

\begin{equation}
D(L)=\frac{n_1n_2L}{(n_1+n_2)^2}\left(1+\frac{L}{n_1+n_2}\right)=\frac{L}{4}\left[1+\frac{2\mu
L}{N(1-2\mu )}\right]\left[1-\frac{4\nu^2}{(1-2\mu )^2}\right]
\label{32}
\end{equation}
and
\begin{equation}
k_0=\frac{n_1L}{n_1+n_2}=\frac{L}{2}\left[1-\frac{2\nu}{1-2\mu}\right].
\end{equation}

In the case of strong persistence (\ref{24a}), the
asymptotic expression for the distribution function
Eq.~(\ref{W(L)}) can be written as
\begin{equation} \label{45f}
W_{L}(k)=W_{L}(0)\frac{n_1L}{k(L-k)}, \,\,\, k\neq 0,\,\, k\neq L,
\end{equation}
\begin{equation}
W_{L}(0)=\frac{n_2}{n_1+n_2} ( 1 - n_1 \ln L ),
\,W_{L}(L)=\frac{n_1}{n_1+n_2} ( 1 - n_2 \ln L ). \label{45b2}
\end{equation}

Both the distribution $W_{L}(k)$ (\ref{45f}) and the function
$W_{N}(k)$ (\ref{21}) have concave forms. The former assumes the
maximal values (\ref{45b2}) at the edges of the interval $[0, L]$
and has a minimum at $k=L/2$.

\subsubsection{Variance $D(L)$}

Using the definition Eq.~(\ref{7}) and the distribution function
Eq.~(\ref{W(L)}) one can obtain a very simple formula for the
variance $D(L)$,

\begin{equation}\label{D(L)}
D(L)=\frac{Ln_1n_2}{(n_1+n_2)^2}\left[1+\frac{(L-1)}{1+n_1+n_2}\right]=\frac{L}{4}\left[1+\frac{2\mu(L-1)}{N-2\mu(N-1)}\right]\left[1-\frac{4\nu^2}
{(1-2\mu)^2}\right].
\end{equation}
Eq.~(\ref{D(L)}) shows that the variance $D(L)$ obeys the
parabolic law independently of the correlation strength in the
Markov chain.

In the case of weak persistence, at $n_2\gg 1$, we obtain the
asymptotics of Eq.~(\ref{32}). It allows one to analyze the behavior
of the variance $D(L)$ with an increase in the ``diffusion time''
$L$. At small $\mu$, the dependence $D(L)$ follows the classical law
of the Brownian diffusion, $D(L)\approx L(1/4-\nu^2)$.

For the case of strong persistence, $n_2 \ll 1$, Eq.~(\ref{D(L)})
gives the asymptotics,
\begin{equation}
D(L)=\frac{n_1n_2L^2}{(n_1+n_2)^2} - \frac{n_1n_2L(L-1)}{n_1+n_2}.
\label{45c}
\end{equation}
The ballistic regime of diffusion leads to the quadratic law of
the $D(L)$ dependence in the zero approximation in the parameter
$n_2 \ll 1$.

The unusual behavior of the variance $D(L)$ raises an issue
as to what particular type of the diffusion equation
corresponds to the nonlinear dependence $D(L)$ in
Eq.~(\ref{32}). In the following subsection, when solving
this problem, we will obtain the conditional probability
$p^{(0)}$ of occurring the symbol zero after a given
$L$-word with $L<N$. The ability to find $p^{(0)}$, with
some reduced information about the preceding symbols being
available, is very important for the study of the
self-similarity of the Markov chain (see Subsubsec.~4 of
this Subsection).

\subsubsection{Generalized diffusion equation at $L<N$, $n_2 \gg 1$}

It is quite obvious that the distribution $W_{L}(k)$ satisfies the
equation
\begin{equation}
W_{L+1}(k)=W_{L}(k)p^{(0)}(k)+W_{L}(k-1)p^{(1)}(k-1).  \label{33}
\end{equation}
Here $p^{(0)}(k)$ is the probability of occurring ''0'' after an
average-statistical $L$-word containing $k$ unities and
$p^{(1)}(k-1)$ is the probability of occurring ''1'' after an
$L$-word containing $(k-1)$ unities. At $L<N$, the probability
$p^{(0)}(k)$ can be written as
\begin{equation}
p^{(0)}(k)=\frac{1}{W_L(k)}
\sum\limits_{i=k}^{k+N-L}p_{i}b(i)\mathrm{C}_{L}^{k}\mathrm{C}_{N-L}^{i-k}.
\label{34}
\end{equation}
The product $b(i)\mathrm{C}_{L}^{k}\mathrm{C}_{N-L}^{i-k}$ in this
formula represents the conditional probability of occurring the
$N$-word containing $i$ unities, the right-hand part of which, the
$L$-sub-word, contains $k$ unities (compare with Eqs.~(\ref{29}),
(\ref{29b})).

The product $b(i)\mathrm{C}_{N-L}^{i-k}$ in Eq.~(\ref{34}) is a
sharp function of $i$ with a maximum at some point $i=i_0$ whereas
$p_{i}$ obeys the linear law (\ref{14}). This implies that $p_{i}$
can be factored out of the summation sign being taken at point
$i=i_0$. The asymptotical calculation shows that point $i_0$ is
described by the equation,

\begin{equation}
i_{0}=\frac{N}{2}\left(1-\frac{2\nu}{1-2\mu}\right)-\frac{L/2}{1-2\mu
(1-L/N)}\left(
1-\frac{2k}{L}-\frac{2\nu}{1-2\mu}\right).\label{35}
\end{equation}
Expression (\ref{14}) taken at point $i_0$ gives
the desired formula for $p^{(0)}$ because
\begin{equation}
\sum\limits_{i=k}^{k+N-L}b(i)\mathrm{C}_{L}^{k}\mathrm{C}_{N-L}^{i-k}
\end{equation}
is obviously equal to $W_L(k)$. Thus, we have

\begin{equation}
p^{(0)}(k)=\frac{1}{2}\left(1+\frac{2\nu}{1-2\mu}\right)+\frac{\mu
L}{N-2\mu (N-L)}\left( 1-\frac{2k}{L}-\frac{2\nu}{1-2\mu}\right).
\label{36}
\end{equation}

Let us consider a very important point relating to Eq.~(\ref{35}).
If the concentration of unities in the right-hand part of the word
(\ref{29b}) is higher than $1/2-\nu/(1-2\mu)$, $k/L
>1/2-\nu/(1-2\mu)$, then the most probable concentration
$(i_0-k)/(N-L)$ of unities in the left-hand part of this word is
likewise increased, $(i_0-k)/(N-L)>1/2-\nu/(1-2\mu)$. At the same
time, the concentration $(i_0-k)/(N-L)$ is less than $k/L$,
\begin{equation}\label{36b}
\frac{1}{2}\left(1-\frac{2\nu}{1-2\mu}\right)
<\frac{i_0-k}{N-L}<\frac{k}{L}.
\end{equation}
This implies that the increased concentration of unities in the
$L$-words is necessarily accompanied by the existence of a certain
tail with an increased concentration of unities as well. Such a
phenomenon is referred by us as the \textit{macro-persistence}. An
analysis performed in the following section will indicate that the
correlation length $l_c$ of this tail is $\gamma N $ with $\gamma
\geq 1$ dependent on the parameters $\mu$ and $\nu$ only. It is
evident from the above-mentioned property of the isotropy of the
Markov chain that there are two correlation tails from both sides of
the $L$-word.

Note that the distribution $W_L(k)$ is a smooth function of
arguments $k$ and $L$ near its maximum in the case of weak
persistence and $k, L-k, Ln_1/n_2\gg 1$. By going over to the
continuous limit in Eq.~(\ref{33}) and using Eq.~(\ref{36}) with
the relation $p^{(1)}(k-1)=1-p^{(0)}(k-1)$, we obtain the
diffusion Fokker-Planck equation for the correlated Markov
process,

\begin{equation}
\frac{\partial W}{\partial L}=\frac{1}{8}\frac{\partial
^{2}W}{\partial \kappa
^{2}}\left(1-\frac{4\nu^2}{(1-2\mu)^2}\right)-
\frac{2\mu}{(1-2\mu)N+2\mu L}\frac{\partial }{\partial \kappa }(
\kappa W), \label{39}
\end{equation}
where $\kappa =k-L/2$. Equation (\ref{39}) has a solution of the
Gaussian form Eq.~(\ref{31}) with the variance $D(L)$ satisfying
the ordinary differential equation,
\begin{equation}
\frac{\mathrm{d}D}{\mathrm{d}L}=\frac{1}{4}\left(1-\frac{4\nu^2}{
(1-2\mu)^2}\right)+\frac{4\mu}{(1-2\mu)N+2\mu L}D. \label{40}
\end{equation}
Its solution, given the boundary condition $D(0)=0$,
coincides with (\ref {32}).

\subsubsection{Self-similarity of the persistent Brownian
diffusion}

In this subsection, we point to one of the most interesting
properties of the Markov chain being considered, namely, its
self-similarity. Let us reduce the $N$-step Markov sequence
by regularly (or randomly) removing some symbols and
introduce the decimation parameter $\lambda$,
\begin{equation}
\lambda =N^{\ast }/N \leq 1.  \label{41}
\end{equation}
Here $N^{\ast }$ is a renormalized memory length for the reduced
$N^{\ast }$-step Markov chain. According to Eq.~(\ref{36}), the
conditional probability $p_{k}^{\ast }$ of occurring the symbol
zero after $k$ unities among the preceding $N^{\ast }$ symbols is
described by the formula,
\begin{equation}
p_{k}^{\ast }=\frac{1}{2}+\nu^\ast+\mu ^{\ast }\left(
1-\frac{2k}{N^{\ast }}\right), \label{42}
\end{equation}
with

\begin{equation}
N^{\ast}=\lambda N, \, \,
\nu^{\ast}=\nu\frac{1}{1-2\mu(1-\lambda)}, \, \, \mu ^{\ast }=\mu
\frac{\lambda }{1-2\mu (1-\lambda )}. \label{43}
\end{equation}
The comparison between Eqs.~(\ref{14}) and (\ref{42}) shows that
the reduced chain possesses the same statistical properties as the
initial one but it is characterized by the renormalized parameters
($N^{\ast }$, $\nu^{\ast}$, $\mu ^{\ast }$) instead of ($N$,
$\nu$, $\mu $). Thus, Eqs.~(\ref{41}) and (\ref{43}) determine the
one-parametrical renormalization of the parameters of the
stochastic process defined by Eq.~(\ref{14}).

The astonishing property of the reduced sequence consists in that
\textit{the variance $D^{\ast }(L)$ is invariant with respect to
the one-parametric decimation transformation} (\ref{41}),
(\ref{43}). In other words, it coincides with the function $D(L)$
for the initial Markov chain:
\begin{equation} \label{44}
D^{\ast }(L) =
\frac{Ln_1^{\ast}}{(n_1^{\ast}+n_2^{\ast})}\left[1+\frac
{(L-1)}{1+n_1^{\ast}+n_2^{\ast}}\right] = D(L), \\ L<N^{\ast }.
\end{equation}
Indeed, according to Eqs.~(\ref{41}), (\ref{43}), the renormalized
parameters $n_1 ^{\ast }=N^{\ast}(1-2(\mu ^{\ast}+\nu^{\ast}))/
4\mu ^{\ast} $ and $n_2 ^{\ast }=N^{\ast}(1-2(\mu
^{\ast}-\nu^{\ast}))/ 4\mu ^{\ast} $ of the reduced sequence
coincides exactly with the parameters $n_1$ and $n_2$ of the
initial Markov chain. Since the shape of the function $W_L(k)$
Eq.~(\ref{W(L)}) is defined by the invariant parameters
$n_1=n_1^{\ast}$ and $n_2=n_2^{\ast}$, the distribution $W_L(k)$
is also invariant with respect to the decimation transformation.

The transformation ($N$, $\nu$, $\mu $) $\rightarrow$ ($N^{\ast
}$, $\nu^{\ast}$, $\mu ^{\ast }$) (\ref{41}), (\ref{43}) possesses
the properties of semi-group, i.\ e., the composition of
transformations ($N$, $\nu$, $\mu $) $\rightarrow$ ($N^{\ast }$,
$\nu^{\ast}$, $\mu ^{\ast }$)  and ($N^{\ast }$, $\nu^{\ast}$,
$\mu ^{\ast }$) $\rightarrow$ ($N^{\ast \ast }$, $\nu^{\ast
\ast}$, $\mu ^{\ast \ast}$) with transformation parameters
$\lambda_1$ and $\lambda_2$ is likewise the transformation from
the same semi-group, ($N$, $\nu$, $\mu$) $\rightarrow$ ($N^{\ast
\ast }$, $\nu^{\ast \ast}$, $\mu ^{\ast \ast}$), with parameter
$\lambda = \lambda_1 \lambda_2$.

\begin{figure}[h!]
{\includegraphics[width=0.45\textwidth,height=0.35\textwidth]{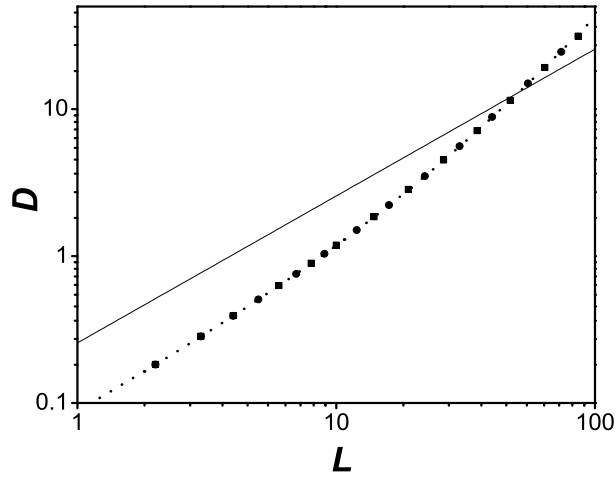}}
\caption{The dependence of the variance $D$ on the tuple length
$L$ for the generated sequence with $N=100$, $\mu=0.4$ and
$\nu=0.08$ (dotted line) and for the decimated sequences (the
parameter of decimation $\lambda =0.5$). Squares and circles
correspond to the stochastic and deterministic reduction,
respectively. The solid line describes the non-correlated Brownian
diffusion, $D(L)=L(1/4-\nu^2)$.} \label{Fig3}
\end{figure}

The invariance of the function $D(L)$ at $L<N$ was referred to by
us as the phenomenon of \textit{self-similarity}. It is
demonstrated in Fig.~\ref{Fig3}.

It is interesting to note that the property of self-similarity is
valid for any strength of the persistency. Indeed, the result
Eq.~(\ref{36}) can be obtained directly from
Eqs.~(\ref{b})-(\ref{17a}), and (\ref{34}) not only for $n_2\gg 1$
but also for the arbitrary value of $n_2$.
\protect
\section{Memory function and its connection with correlation function}

Typically, the correlation function and other moments are employed
as the input characteristics for the description of the correlated
random sequences. However, the correlation function describes not
only the direct interconnection of the elements $a_i$ and
$a_{i+r}$, but also takes into account their indirect interaction
via all other intermediate elements. Our approach operates with
the "origin" characteristics of the system, specifically, with the
memory function. The correlation and memory functions are
mutual-complementary characteristics of a random sequence in the
following sense. The numerical analysis of a given random sequence
enables one to directly determine the correlation function rather
than the memory function. On the other hand, it is possible to
construct a random sequence using the memory function, but not the
correlation one. Therefore, we believe that the investigation of
memory function of the correlated systems will permit one to
disclose their intrinsic properties which provide the correlations
between the elements.

The memory function used in Refs.~\cite{uyakm,uya} was
characterized by the step-like behavior and defined by two
parameters only: the memory depth $N$ and the strength of symbol's
correlations. Such a memory function describes only one type of
correlations in a given system, the persistent or anti-persistent
one, which results in the super- or sub-linear dependence
$D(L)$~\cite{rem}. Obviously, both types of correlations can be
observed at different scales in the same system. Thus, one needs
to use more complex memory functions for detailed description of
the systems with both type of correlations. Besides, we have to
find out a relation connecting the mutually-complementary
characteristics of random sequence, the memory and correlation
functions.

\subsection{Main equation}

Let us rewrite Eq.~(\ref{1}) in an equivalent form,
\begin{equation}
P(a_{i}=1\mid T_{N,i})=b+\sum\limits_{r=1}^{N} F(r)(a_{i-r}-b),
\label{m2a}
\end{equation}
with
\begin{equation}\label{m2c}
b=\frac{\sum\limits_{r=1}^{N}f(0,r)/N}{1-\sum\limits_{r=1}^{N}
F(r)}, \qquad F(r)=\frac{1}{N}[f(1,r)-f(0,r)].
\end{equation}
The constant $b$ is the value of $a_{i}$ averaged over the whole
sequence, $b=\bar{a}$:
\begin{equation}\label{m2e}
\bar{a}=\lim_{M\rightarrow
\infty}\frac{1}{2M+1}\sum_{i=-M}^{M}a_i.
\end{equation}
Indeed, according to the ergodicity of the Markov chain, $\bar{a}$
coincides with the value of $a_i$ averaged over the ensemble of
realizations of the Markov chain. So, we can write
\begin{equation}
\label{m2b} \bar{a}=Pr(a_{i}=1)=\sum\limits_{T_{N,i}}P(a_{i}=1\mid
T_{N,i})Pr(T_{N,i}).
\end{equation}
Here $Pr(a_{i}=1)$ is the probability of occurring the symbol
$a_{i}$ equal to unity and $Pr(T_{N,i})$ is the probability of
occurring the definite word $T_{N,i}$ in the considering ensemble
of sequences. Substituting $P(a_{i}=1\mid T_{N,i})$ from
Eq.~(\ref{m2a}) into Eq.~(\ref{m2b}) and taking into account the
obvious relation $\sum\limits_{T_{N,i}}Pr(T_{N,i})=1$, one gets,
\begin{equation}\label{m2d}
\bar{a}=b-b\sum\limits_{r=1}^{N}F(r) + \sum\limits_{r=1}^{N}F(r)
\sum\limits_{T_{N,i}}Pr(T_{N,i})a_{i-r}.
\end{equation}
The sum $\sum\limits_{T_{N,i}}Pr(T_{N,i})a_{i-r}$ does not depend
on the subscript $r$ and obviously coincides with $\bar{a}$. So,
we have $\bar{a}=b+(\bar{a}-b)\sum\limits_{r}F(r)$. From this
equation we conclude that $b=\bar{a}$. Thus, we can rewrite
Eq.~(\ref{m2a}) as
\begin{equation}\label{m2}
P(a_{i}=1 \mid T_{N,i})=\bar{a}+\sum\limits_{r=1}^{N}
F(r)(a_{i-r}-\bar{a}).
\end{equation}

We refer to $F(r)$ as the \emph{memory function} (MF). It
describes the strength of influence of previous symbol $a_{i-r}$
upon a generated one, $a_{i}$. To the best of our knowledge, the
concept of memory function for many-step Markov chains was
introduced in Ref.~\cite{uyakm}. The function $P(. \mid .)$
contains the complete information about correlation properties of
the Markov chain.

We suggest below two methods for finding the memory function
$F(r)$ of a random binary sequence with a known correlation
function. The first one is based on the minimization of a
"distance" $\emph{Dist}$ between the Markov chain generated by
means of a sought-for MF and the initial sequence of symbols. This
distance is determined by the formula,
\begin{equation}\label{optim1}
\emph{Dist}=\overline{(a_{i}-P(a_{i}=1 \mid T_{N,i}))^2} =
\lim_{M\rightarrow
\infty}\frac{1}{2M+1}\sum_{i=-M}^{M}(a_{i}-P(a_{i}=1 \mid
T_{N,i}))^2,
\end{equation}
with the conditional probability $P$ defined by Eq.~(\ref{m2}).

Let us express distance~(\ref{optim1}) in terms of the correlation
function,
\begin{equation}\label{cor}
K(r)=\overline{a_{i}a_{i+r}}-\bar{a}^{2},\  \
K(0)=\bar{a}(1-\bar{a}),\  \ K(-r)=K(r).
\end{equation}
From Eqs.~(\ref{m2}), (\ref{optim1}), one obtains
\[
\emph{Dist}=\sum\limits_{r,r'}\overline{(a_{i-r}-\bar{a})(a_{i-r'}
-\bar{a})}F(r)F(r')
-2\sum\limits_{r}\overline{(a_{i}-\bar{a})(a_{i-r}-\bar{a})}F(r)
+\overline{(a_{i}-\bar{a})^{2}}
\]
\begin{equation}
=\sum\limits_{r,r'}K(r-r')F(r)F(r')-2\sum\limits_{r}K(r)
F(r)+K(0). \label{optim3}
\end{equation}
The minimization equation,
\begin{equation}
\frac{\delta\emph{Dist}}{\delta
F(r)}=2\sum\limits_{r'}K(r-r')F(r')-2K(r)=0, \label{optim4}
\end{equation}
yields the relationship between the correlation and memory
functions,
\begin{equation} \label{mmain}
K(r)=\sum\limits_{r'=1}^{N}F(r')K(r-r'), \ \ \ \ r\geq 1.
\end{equation}
Equation~(\ref{mmain}) can also be derived by straightforward
calculation of the average $\overline{a_{i}a_{i+r}}$ in
Eq.~(\ref{cor}) using definition~(\ref{m2}) of the memory
function. \protect\begin{figure}[h!]
\begin{centering}
\scalebox{0.8}[0.8]{\includegraphics{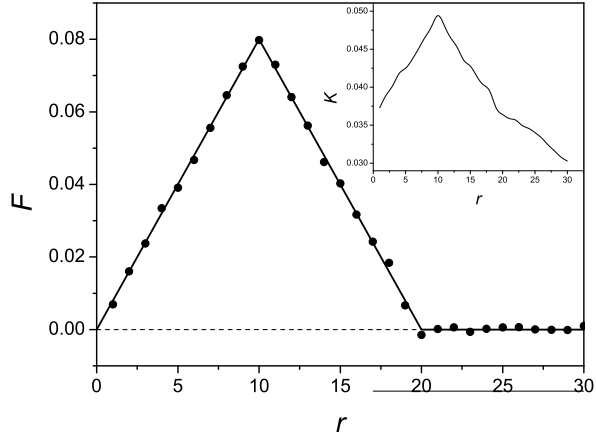}} \caption{The
initial memory function Eq.~(\ref{eqmf}) (solid line) and the
reconstructed one (dots) vs the distance $r$. In inset, the
correlation function $K(r)$ obtained by a numerical analysis of
the sequence constructed by means of the memory function
Eq.~(\ref{eqmf}).} \label{Fig11}
\end{centering}
\end{figure}

The second method resulting from the first one, establishes a
relationship between the memory function $F(r)$ and the variance
$D(L)$,
\begin{equation}
\label{MF} M(r,0)=\sum\limits_{r'=1}^{N}F(r')M(r,r'),
\end{equation}
\[
M(r,r')=D(r-r')-(D(-r')+r[D(-r'+1)-D(-r')]).
\]
It is a set of linear equations for $F(r)$ with coefficients
$M(r,r')$ determined by $D(r)$. The relations,
$K(r)=[D(r-1)-2D(r)+D(r+1)]/2$ obtained in Ref.~\cite{uyakm} and
$D(-r)=D(r)$ are used here.

\protect\begin{figure}[h!]
\begin{centering}
\scalebox{0.8}[0.8]{\includegraphics{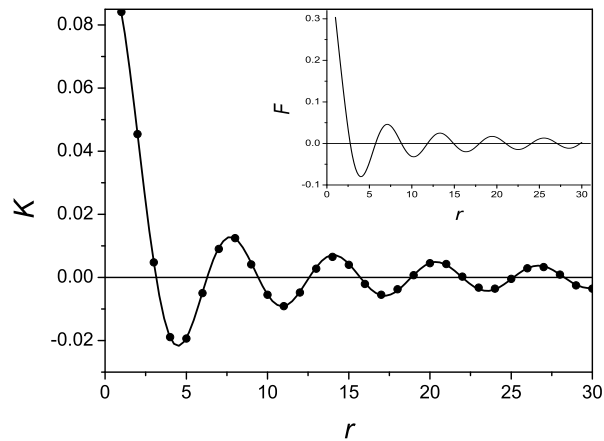}} \caption{The
model correlation function $K(r)$ described by Eq.~(\ref{K(r)})
(solid line). The dots correspond to the reconstructed correlation
function. In inset, the memory function $F(r)$ obtained by
numerical solution of Eq.~(\ref{mmain}) with correlation function
Eq.~(\ref{K(r)}).} \label{Fig12}
\end{centering}
\end{figure}

Let us verify the robustness of our method by numerical
simulations. We consider a model "triangle" \emph{memory
function},
\begin{equation}
F(r)=0.008\cases {r,\;\qquad \;\;\ 1 \leq r < 10,  \cr 20-r,\;\;\
10 \leq r < 20 ,\cr 0, \;\;\;\;\;\;\;\;\;\;\ r\geq 20,}
\label{eqmf}
\end{equation}
presented in Fig.~\ref{Fig11} by solid line. Using Eq.~(\ref{m2}),
we construct a random non-biased, $\bar{a}=1/2$, sequence of
symbols $\{0,1\}$. Then, with the aid of the constructed binary
sequence of the length $10^6$, we calculate numerically the
correlation function $K(r)$. The result of these calculations is
presented in inset Fig.~\ref{Fig11}. One can see that the
correlation function $K(r)$ mimics roughly the memory function
$F(r)$ over the region $1\leq r \leq 20$. In the region $r>20$,
the memory function is equal to zero but the correlation function
does not vanish~\cite{ref}. Then, using the obtained correlation
function $K(r)$, we solve numerically Eq.~(\ref{mmain}). The
result is shown in Fig.~\ref{Fig11} by dots. One can see a good
agrement of initial, Eq.~(\ref{eqmf}), and reconstructed memory
functions $F(r)$.

\subsection{Numerical simulations}

The main and very nontrivial result of our paper consists in the
ability to construct a binary sequence with an arbitrary
\emph{prescribed correlation function} by means of
Eq.~(\ref{mmain}). As an example, let us consider the model
correlation function,
\begin{equation}
K(r) = 0.1 \frac{\sin(r)}{r}, \label{K(r)}
\end{equation}
presented by the solid line in Fig.~\ref{Fig12}. We solve
Eq.~(\ref{mmain}) numerically to find the memory function $F(r)$
using this correlation function. The result is presented in inset
Fig.~\ref{Fig12}. Then we construct the binary Markov chain using
the obtained memory function $F(r)$. To check up a robustness of
the method, we calculate the correlation function $K(r)$ of the
constructed chain (the dots in Fig.~\ref{Fig12}) and compare it
with Eq.~(\ref{K(r)}). One can see an excellent agreement between
the initial and reconstructed correlation functions.

\protect\begin{figure}[h!]
\begin{centering}
\scalebox{0.8}[0.8]{\includegraphics{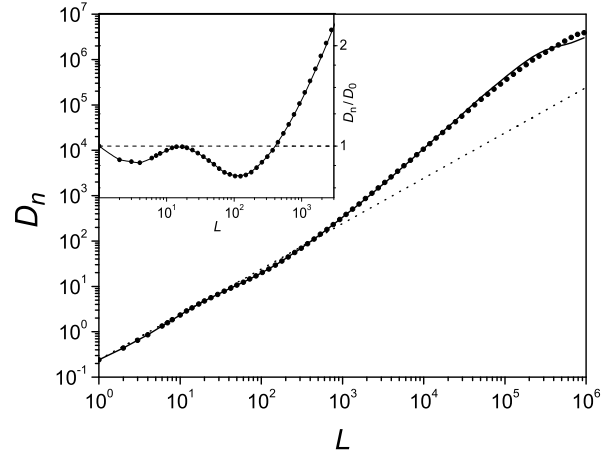}} \caption{The
normalized variance $D_n(L)$ for the coarse-grained text of Bible
(solid line) and for the sequence generated by means of the
reconstructed memory function $F(r)$ (dots). The dotted straight
line describes the non-biased non-correlated Brownian diffusion,
$D_{0}(L)=L/4$. The inset demonstrates the anti-persistent
dependence of ratio $D_n(L)/D_0(L)$ on $L$ at short distances.}
\label{Fig13}
\end{centering}
\end{figure}
Let us demonstrate the effectiveness of our concept of the
additive Markov chains when investigating the correlation
properties of coarse grained literary texts. First, we use the
coarse-graining procedure and map the letters of the text of
Bible~\cite{bibe} onto the symbols zero and unity (here, $(a-m)
\mapsto 0, (n-z) \mapsto 1$). Then we examine the correlation
properties of the constructed sequence and calculate numerically
the variance $D(L)$. The result of simulation of the normalized
variance $D_n(L)=D(L)/4 \bar{a}(1-\bar{a})$ is presented by the
solid line in Fig.~\ref{Fig13}. The dominator
$4\bar{a}(1-\bar{a})$ in the equation for the normalized variance
$D_n(L)$ is inserted in order to take into account the inequality
of the numbers of zeros and unities in the coarse-grained literary
texts. The straight dotted line in this figure describes the
variance $D_{0}(L)=L/4$, which corresponds to the
\textit{non-biased non-correlated} Brownian diffusion. The
deviation of the solid line from the dotted one demonstrates the
existence of correlations in the text. It is clearly seen that the
diffusion is anti-persistent at small distances, $L\alt 300$, (see
inset Fig.~\ref{Fig13}) whereas it is persistent at long
distances.

\begin{figure}[h!]
\begin{centering}
\scalebox{0.8}[0.8]{\includegraphics{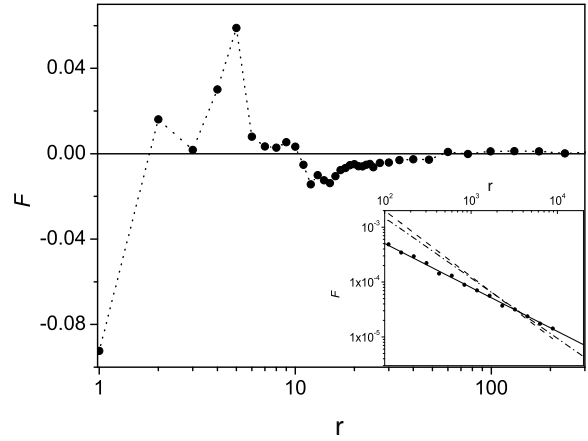}} \caption{The
memory function $F(r)$ for the coarse-grained text of Bible at
short distances. In inset, the power-law decreasing portions of
the $F(r)$ plots for several texts. The dots correspond to
"Pygmalion" by B. Shaw. The solid line corresponds to power-law
fitting of this function. The dash dotted and dashed lines
correspond to Bible in English and Russian, respectively.}
\label{Fig14}
\end{centering}
\end{figure}

The memory function $F(r)$ for the coarse-grained text of Bible at
$r<300$ obtained by numerical solution of Eq.~(\ref{MF}) is shown
in Fig.~\ref{Fig14}. At long distances, $r>300$, the memory
function can be nicely approximated by the power function
$F(r)=0.25r^{-1.1}$, which is presented by the dash-dotted line in
inset Fig.~\ref{Fig14}.

\protect

Note that the region $r \alt 40$ of negative anti-persistent
memory function provides much longer distances $L \sim 300$ of
anti-persistent behavior of the variance $D(L)$.

Our study reveals the existence of two characteristic regions with
different behavior of the memory function and, correspondingly, of
persistent and anti-persistent portions in the $D(L)$ dependence.
This appears to be a prominent feature of all texts written in any
language. The positive persistent portions of the memory functions
are given in inset Fig.~\ref{Fig14} for the coarse-grained
English- and Russian-worded texts of Bible (dash-dotted and dashed
lines, Refs.~\cite{bibe} and~\cite{bibr}, correspondingly).
Besides, for comparison, the memory function of the coarse-grained
text of "Pygmalion" by B. Shaw~\cite{pyg} is presented in the same
inset (dots), the power-law fitting is shown by solid line.

\begin{figure}[h!]
\begin{centering}
\scalebox{0.8}[0.8]{\includegraphics{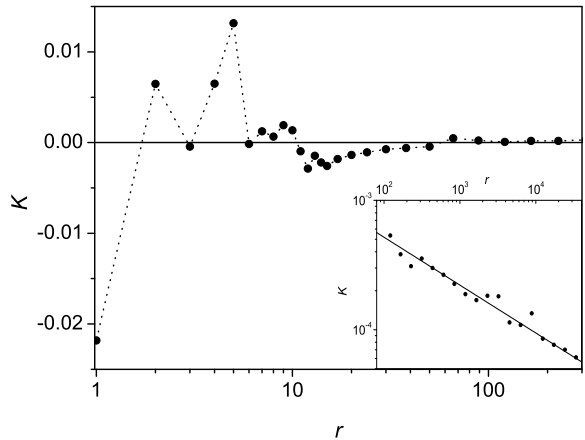}} \caption{The
correlation function $K(r)$ for the coarse-grained text of Bible
at short distances. In inset, the power-law decreasing portions of
the $K(r)$ plot for the same text. The solid line corresponds to
power-law fitting of this function.} \label{Fig15}
\end{centering}
\end{figure}

It is interesting to note that the memory function of any text
mimics the correlation function, as it was found for the model
example Eq.~(\ref{K(r)}). This fact is confirmed by
Fig.~\ref{Fig15} where the correlation function of the
coarse-grained text of Bible is shown. One can see that its
behavior at both short and long scales is similar to the memory
function presented in Fig.~\ref{Fig14}. However, the exponents in
the power-law approximations of $K(r)$ and $F(r)$ functions differ
essentially.

\protect

\section{Conclusion}

Thus, the simple, exactly solvable model of the uniform binary
$N$-step Markov chain is presented. The memory length $N$, the
parameter $\mu$ of the persistent correlations and the biased
parameter $\nu$ are three parameters in our theory. The
correlation function ${\cal K}(r)$ is usually employed as the
input characteristics for the description of the correlated random
systems. Yet, the function ${\cal K}(r)$ describes not only the
direct interconnection of the elements $a_i$ and $a_{i+r}$, but
also takes into account their indirect interaction via other
elements. Since our approach operates with the ''original''
parameters $N$, $\mu$ and $\nu$, we believe that it allows us to
reveal the intrinsic properties of the system which provide the
correlations between the elements.

We have demonstrated the efficiency of description of the symbolic
sequences with long-range correlations in terms of the memory
function. An equation connecting the memory and correlation
functions of the system under study is obtained. This equation
allows reconstructing a memory function using a correlation function
of the system. Actually, the memory function appears to be a
suitable informative "visiting card" of any symbolic stochastic
process. The effectiveness and robustness of the proposed method is
demonstrated by simple model examples. Memory functions for some
concrete examples of the coarse-grained literary texts are
constructed and their power-law behavior at long distances is
revealed. Thus, we have shown the complexity of organization of the
literary texts in contrast to a previously discussed simple
power-law decrease of correlations~\cite{kant}.

If the memory length $N$ of the system under consideration is of
order of the very system length then the Markov chain, modeling the
system, could be non-stationary. In this case the proposed method
does not allow to describe the system precisely, as distinct from
the method proposed in~\cite{Grig1,Grig2}.

\section{Appendix. Matrix of the conditional probability}

In this Appendix, we prove the property of metrical transitivity
of the $N$-step Markov chains.

It is possible to look at the Markov chain from the other point of
view and consider it as a $1$-step \emph{vector} Markov chain. To
this end, we introduce the $N$-component vector-function $X_l$,
\begin{equation}\label{eq5}
X_l=(a_{l+1},a_{l+2},...,a_{l+N}), \qquad l=...,-2,-1,0,1,2,...
\end{equation}
The number of different sets of symbols
$(a_{i+1},a_{i+2},...,a_{i+N})$ is equal to $Q=2^N$. We number the
different states of the vector $X_l$ by their binary
representation,
\begin{equation}\label{eq51}
D(a_{N},a_{N-1},...,a_1)=a_N2^0+a_{N-1}2^1+...+a_12^{N-1}, \qquad
0\leq D \leq 2^N-1.
\end{equation}
The matrix elements $M_{ik}$ of the probability matrix $M$, i.e.
the probabilities of transition of the vector
$X=(a_1,a_2,...,a_N)$ into the vector $Y=(a_1',a_2',...,a_N')$ can
be expressed via the function of conditional probability
$P(a_{i}\mid T_{N,i})$. The subscripts $i$ and $k$ of the matrix
$M_{ik}$ are determined by the binary representations of the
sequences $(a_1,a_2,...,a_N)$ and $(a_1',a_2',...,a_N')$,
correspondingly: $i=1+D(a_{N},a_{N-1},...,a_1)$ and
$k=1+D(a_{N}',a_{N-1}',...,a_1')$. Every matrix row contains only
two non-zero elements since the vector $X_1$ can take up two
values only, namely, $(a_2,a_3,...,a_N,0)$ and
$(a_2,a_3,...,a_N,1)$. For $k \leq Q/2$, let us denote the
probability of occurring of $a_{N+1}=0$ as $1-P_k$, where the
index $k$ is equal to $k=1+D(a_{N},a_{N-1},...,a_1)$ in the binary
representation.

For the index $k$ being in the range from $Q/2+1$ to $Q$, we
denote the probability of occurring of symbol $a_{N+1}=0$ after
the word $a_1,a_2,...,a_N$ as $P_k$. Then, $1-P_k$ is the
probability of occurring of the symbol unity. Taking into account
that $a_N=0$ for $k\leq Q/2$ and obvious relations,
\[
D(a_{N-1},...,a_1,0)=2D(a_{N},a_{N-1},...,a_1),
\]
\[
D(a_{N-1},...,a_1,1)=2D(a_{N},a_{N-1},...,a_1)+1,
\]
we get the transition probabilities matrix $M$:
\begin{equation}\label{eq7}
M=\left(%
\begin{array}{cccccccc}
  1-P_1 & P_1 & 0 & 0 & ... & ... & 0 & 0 \\
  0 & 0 & 1-P_2 & P_2 & 0 & ... & 0 & 0 \\
  ... & ... & ... & ... & ... & ... & ... & ... \\
  0 & 0 & ... & ... & 0 & 0 & 1-P_{Q/2} & P_{Q/2} \\
  1-P_{Q/2+1} & P_{Q/2+1} & 0 & 0 & ... & ... & 0 & 0 \\
  0 & 0 & 1-P_{Q/2+2} & P_{Q/2+2} & 0 & ... & 0 & 0 \\
  ... & ... & ... & ... & ... & ... & ... & ... \\
  0 & 0 & ... & ... & 0 & 0 & 1-P_Q & P_Q \\
\end{array}%
\right).
\end{equation}
Thus, to determine the vectors $b$ of probability distribution of
$N$-words in the stationary Markov chain we need to solve the
system of equations,
\begin{equation}\label{eq8}
b_i=\sum\limits_{k=1}^{Q}b_kM_{ki}, \qquad\qquad\qquad
\sum\limits_{k=1}^{Q}b_k=1.
\end{equation}
In other words, one needs to obtain the normalized eigenvector
corresponding to the eigenvalue $\lambda = 1$ of the matrix
$M^{(N)}$ of the order $Q=2^N$. It is clear that if the vector $b$
satisfies to the condition $bM=b$ then for every integer $k$ the
condition $bM^k=b$ is also true, here $M^k$ is the power $k$ of
the matrix $M$. Let us consider the matrix $M^N$ and show that all
matrix elements are positive. In this case, following the Markov
theorem we can conclude that the matrix $M$ determines uniquely
the probability of the words distribution.

Let us suggest that for any $k < N$ the matrix $M^k$ satisfies to
the next conditions: in the first row the elements $M_{1i}$ for
$i=1,\ldots,2^k$ are positive, in the second row the positive
elements are $M_{2i}$ with $i=2^k+1, \ldots 2\times 2^k$, ... in
the $2^{N-k}$-th row --- $i=(2^{N-k}-1)2^k, \ldots, 2^N$. In the
next rows this order is repeated. Let us demonstrate that if the
matrix $M^k$ obeys to this rules, then it is true for the matrix
$M^{k+1}$ also.

After multiplication of matrixes $M^k$ and $M$ the elements of
obtained matrix are defined by the expression:
\begin{equation}\label{eq8a}
M^{k+1}(i,j) = \sum\limits_l M^k(i,l)M(l,j).
\end{equation}
Let us consider the first row of the matrix $M^{k+1}$ --- $i=1$.
In each column of the matrix $M$ only two elements are non-zero.
After multiplication of the first row of the matrix $M^k$ to some
column of the matrix $M$ the result is non-zero (positive) for $j
\leq 2*2^k$ only, because positive elements of the matrix $M$
corresponds to the positive zone ($i < 2^k$) of the first row of
matrix $M^k$ only for this $j$. So the described rule remains for
the first row of the matrix $M^{K+1}$. Similarly this fact can be
proved for other rows.

The matrix $M^1$ obeys to this rule, consequently, by induction,
it is true for all $M^k$. In according to this rule, if power
$k=N$, then all elements of the matrix $M^N$ are positive.

Therefore, from the Markov theorem, there is the unique solution
of the system $bM^N=b$ (or $bM=b$). This solution can be obtained
by the method of successive approximations,
\begin{equation}\label{eq8b}
b^{k+1}_i = b^{k}_j M(j,i),\qquad k=0,1,2,...,
\end{equation}
if we start from the arbitrary initial distribution $b^{0}_j$. In
the limit $k\rightarrow \infty$ we get to the stationary
distribution of the probability vector $b$.

Taking into account the explicit form of the matrix $M$, the
equation~(\ref{eq8}) comes to the next equations:
\begin{equation}\label{eq9}
b_i(1-P_i)+b_{i+Q/2}P_{i+Q/2}=b_{2i-1},
\end{equation}
\[
b_iP_i+b_{i+Q/2}(1-P_{i+Q/2})=b_{2i}.
\]
For $Q=2$ we get the well known result~\cite{gn}:
\[
M =\left(%
\begin{array}{cc}
  1-P_1 & P_1 \\
  P_2 & 1-P_2 \\
\end{array}%
\right),
\]
\[
b_1=\frac{P_2}{P_1+P_2}, \qquad   b_2=\frac{P_1}{P_1+P_2}.
\]
And in the case $Q=4$ we obtain the next result:
\[
M = \left(%
\begin{array}{cccc}
  1-P_1 & P_1 & 0 & 0 \\
  0 & 0 & 1-P_2 & P_2 \\
  P_3 & 1-P_3 & 0 & 0 \\
  0 & 0 & P_4 & 1-P_4 \\
\end{array}%
\right),
\]
\[
b_1=\frac{P_3p_4}{P_1P_2+2P_1P_4+P_3P_4}, \qquad
b_2=b_3=\frac{P_1P_4}{P_1P_2+2P_1P_4+P_3P_4}, \qquad
b_4=\frac{P_1P_2}{P_1P_2+P_1P_4+P_3P_4}.
\]


\begin{references}
\bibitem{rem} Note that we here discuss the dependence of the
variance $D$ upon the length $L$ that describes the persistent (or
antipersistent) correlations in the words of different lengths
$L$. This length does not coincides with the memory length $N$, $L
\neq N$. The dependence $D(L)$ is completely different from the
dependence $D$ on the memory length $N$ discussed in
Ref.~\cite{nar1,nar2}.
\bibitem{ref} The existence of the "additional tail" in the correlation function
is in agreement with Ref.~\cite{uyakm} and corresponds to the well
known fact that the correlation length is always larger then the
region of memory function action.
\end{references}
\end{document}